# An Open-RAN Testbed for Detecting and Mitigating Radio-Access Anomalies

Hanna Bogucka[1,2], Marcin Hoffmann[1,2], Paweł Kryszkiewicz[1,2], Łukasz Kułacz[1,2]

[1]Poznan University of Technology, Poland; [2]RIMEDO Labs, Poland

**Abstract**

This paper presents the Open Radio Access Network (O-RAN) testbed for secure radio access. We discuss radio-originating attack detection and mitigation methods based on anomaly detection and how they can be implemented as specialized applications (xApps) in this testbed. We also present illustrating results of the methods applied in real-world scenarios and implementations.

## I. Introduction

The traditional way of providing the Radio Access Network (RAN) is a black (closed) box supplied by a single vendor. The Open-RAN (O-RAN) concept is based on two ideas: the disaggregation of RAN functionality from the underlying hardware (introduced in the Fifth Generation (5G) radio specifications by 3GPP) and the definition of functional RAN modules with interfaces [1]. The disaggregation and modularity of O-RAN allow for an open ecosystem for vendors, software developers, and system integrators.

In 5G RAN, the functions of the base station are split into entities with open interfaces: a centralized unit (CU) with a separated User-Plane (CU-UP) having N3 interface with User Plane Function (UPF) and Control-Plane (CU-CP) having N2 interface with Access and Mobility Function (AMF), a distributed unit (DU), and a remote radio unit (RU). With the O-RAN approach, different vendors can develop those entities due to the open interfaces, including Open Front-Haul (O-FH). Moreover, the RAN Intelligent Controllers (RICs) is a new component that allows the provision of Artificial Intelligence (AI)-based management for radio network automation.

Figure 1 presents the O-RAN architecture. The white modules (including functionalities and interfaces) are open versions of the 3GPP specified nodes DU, CU-CP, CU-UP, and RU, which are evolved by O-RAN to support open interfaces with new services. In contrast, orange blocks and interfaces are specified by the O-RAN Alliance. The "O-" prefix (in O-CU, O-DU, O-RU) stands for "open." At least one O-CU-CP, one O-CU-UP, and one O-DU comprise the E2 node. RICs are: Non-Real-Time RIC (Non-RT RIC) and Near-Real-Time RIC (Near-RT RIC). Non-RT RIC is a software platform for applications called rApps for high-level RAN optimization in a *non-real-time control loop* (with a latency of one second or more). It allows RAN configuration and management and provides enrichment information and guidance via policies and Machine Learning (ML) models to the Near-RT RIC via the A1 interface. Near-RT RIC is a software platform for a set of xApps, which are applications running in a *near real-time control loop*, i.e., in a time scale longer than ten milliseconds and shorter than one second. It enables near-RT control and optimization of RAN elements and resources via fine-grained data collection and actions over the E2 interface [2]. Service and Management Orchestration (SMO) entity supports the orchestration of O-RAN components, including Non-RT RIC. Finally, O-Cloud is a cloud computing platform comprising a collection of physical infrastructure nodes that meet O-RAN requirements to host the relevant O-RAN functions (Near-RT RIC, O-CU-CP, O-CU-UP, and O-DU), the supporting software components (such as Operating System, Virtual Machine Monitor, Container Runtime, etc.) and the management and orchestration functions. Details of O-RAN blocks and interfaces can be found in [1].

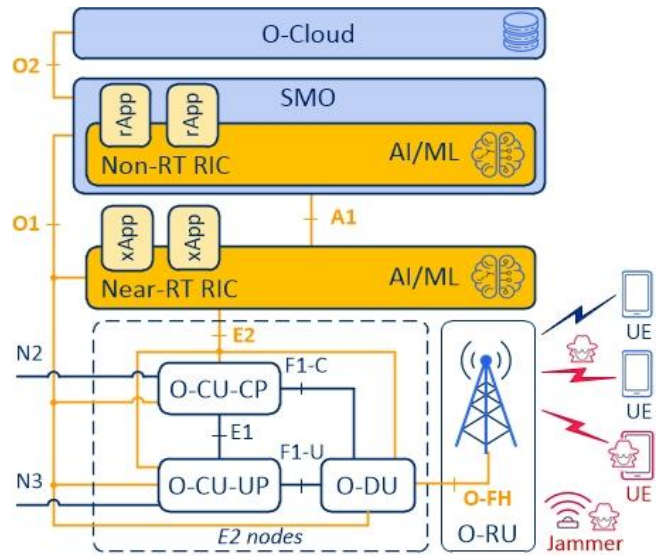

Figure 1. O-RAN architecture and radio attacks.

The O-RAN concept poses a challenge to its security. Inadequately defined and poorly secured O-RAN interfaces and applications may be vulnerable to cyberattacks. Moreover, in a decentralized approach to information processing, security management becomes difficult as significant parts of the network can be attacked anywhere at any time. These security issues result from O-RAN's novel features: *openness* and *intelligence*.

However, these same features of O-RAN create opportunities for detection and response to attacks. The transparency of new open interfaces increases scrutiny of vulnerabilities and failures.

O-RAN security challenges and solutions have been overviewed and discussed in several recent papers (e.g., [2]-[5]), MNOs' documents (e.g., [6]), and governmental statements (e.g., [7]). Importantly, all stakeholders agree that O-RAN brings more competition to implementing security solutions usable with equipment from any O-RAN-compliant vendor. O-RAN RICs provide a framework to enhance security, resilience, and adaptability. Embedded intelligence allows for monitoring and analysis of security threats and protects RAN from malicious and illegal access to network segments. It makes it possible to detect threats before they affect the operation of the entire network. Significantly, xApps and rApps can be developed for specific types of threats in a network, which can be detected and eliminated closer to the place of their occurrence.

This paper discusses the security opportunities the O-RAN architecture brings. We present an O-RAN testbed with xApps for anomaly detection in the radio segment, near-real-time reactions and defense to attacks launched. To our knowledge, the proposed algorithms are pioneers in the field of O-RAN. Unlike the simulation studies, the implementation of such algorithms on real hardware is limited first to what O-RAN specification supports and second to what a particular platform supports in terms of input/output data.

Section II discusses the attack surface in the radio segment, mitigation methods, and O-RAN security opportunities. In Section III, we present our O-RAN testbed implementing xApps for anomaly detection and mitigation methods. Sections IV and V present xApps developed for jamming and signaling storm detection and mitigation, respectively, as well as their measured performance in our testbed implementations. We conclude the paper in Section VI.

## II. O-RAN for radio-access security

Virtualization, functional disaggregation, and open interfaces in O-RAN inevitably extend the threat surface. The Security Focus Group (SFG) working group (WG11) within the O-RAN Alliance has identified six groups of threat surfaces (causes of attack vulnerabilities) for O-RAN [8]. Four of these groups are related to O-RAN's openness, virtualization, disaggregation, and intelligence. The other two are the consequences of accessibility in the supply chain and provided software (SW). Apart from the threat surfaces mentioned, [8] identifies eight threat categories, depending on the goals of attacks. Noticeably, some threats would appear in standard RANs, computer networks, or cloud services, not exclusively in O-RAN. This paper focuses on radio-segment security threats, particularly those detected as anomalies, i.e., causing signals or observations to deviate from standard or expected ones, making them inconsistent with the rest.

*A. Attacks on the radio segment and anomalies*

The radio interface is inherently exposed to attacks related to the omnipresent transmission medium. A standard way to assess risks is based on their threat to *confidentiality*, *integrity*, and *availability* of transferred information.

*Confidentiality* in RAN can be compromised by *eavesdropping*, data decryption and detection capabilities, and traffic analysis. *The integrity* of information is lost when an adversary intercepts and tampers with it in a way that tricks the recipient into behaving differently. *Spoofing* (disguising a communication or user identity) or *Man in the Middle* (MITM) attacks (intercepting messages and changing their content) are typically considered attacks in RANs. Finally, the information *availability* may be compromised by attacks that interrupt connections or exhaust network resources and cause traffic congestion. Threats in this category typically considered in RANs are jamming (which injects unwanted signals into the communication or control channels), Signaling Storm (SS) (seizing resources for signaling in the control plane), or *Denial of Service* (DoS) (aimed at massive requests for network resources, either in the control or user plane). Note that according to MITRE FiGHT™ (5G Hierarchy of Threats) knowledgebase, SS is a subcategory of a DoS attack, i.e., if launched in the random-access channel, its consequence might be denial of access to a network for legitimate users. The literature (e.g., [9]) extensively describes radio attacks and appropriate countermeasures. As shown in Figure 1, they can be launched at UE, in a radio link, or in an external radio source. In what follows, we discuss the O-RAN security testbed and the two common radio attacks: jamming and SS.

Jamming may be of different kinds [10]:
- *regular*, when jammers do not follow any medium access control protocol,
- *delusive*, when jammers inject a legitimate sequence of bits in a communication channel,
- *random*, when jammers conserve their energy by alternating between active and idle states,
- *responsive*, when jammers transmit only if the legitimate transmitter is active,
- *go-next*, when a jammer targets one frequency channel at a time and follows frequency hops

of the legitimate transmitter or
- *control-channel jamming*, when a jammer targets the control message exchanges.

In general, jamming acts like noise, keyed noise, or interference; therefore, it can be detected if it deteriorates the reception quality. However, there are no means to remove it from the useful signal. Mitigating its negative effects involves adapting to the deteriorated radio conditions. Smart (delusive) jamming mimics legitimate transmission (e.g., 5G) and is harder to detect, but it requires specialized equipment and adequate transmission power (as in the victim system). Such a jammer can be treated like a spoofing device.

The latter type of anomaly attack considered, SS, can be prevalent since it is relatively easy to launch using a cheap, authorized device, such as a prepaid phone. In [11], models for SSs and existing detection and mitigation solutions are presented, although this paper focuses on Radio Resource Control (RRC). In general, the methods counteracting the SS attack aim at detecting and breaking suspicious connections or blocking access to the signaling channels.

*B. O-RAN opportunities for anomaly detection/mitigation*

An essential element in introducing O-RAN security is the inclusion of the security subsystem in the Near-RT RIC architecture. Its primary task is to prevent malicious xApps from affecting the O-RAN performance and data leaks. Additionally, mechanisms to mitigate conflicts between xApps should be a security practice for O-RAN.

Regarding attack detection and defense in the radio segment, O-RAN architecture has vast potential, mainly thanks to the specialized xApps and rApps that protect RAN against attacks. They can be easily installed and updated as new threats occur and new solutions and countermeasures are elaborated. Moreover, xApps and rApps address the zero-trust model by continuously monitoring, detecting, and reacting to threats, including anomalies in the radio traffic [12].

### III. O-RAN security testbed and real-world experiments

As said above, O-RAN architecture supports system security through Near-RT RIC and Non-RT RIC and specialized xApps and rApps designed to detect and eliminate threats promptly before they impact the network. Below, we present a testbed to run and evaluate security xApps for ML-based anomaly detection. The example scenarios and xApps are chosen to illustrate attacks, their detection, and mitigation in selected representative scenarios: jamming in Ultra-Reliable, Low Latency Communication (URLLC) and SS in massive Machine-Type Communication (mMTC). The selection of these use cases will be justified in the following sections. Figure 1 presents the general scenario for our setup, and Figure 2 shows the equipment used for monitoring the attacks and mitigation applications. Here, we consider just a one-cell scenario since the detection and mitigation methods presented here operate in a single cell or even a single link.

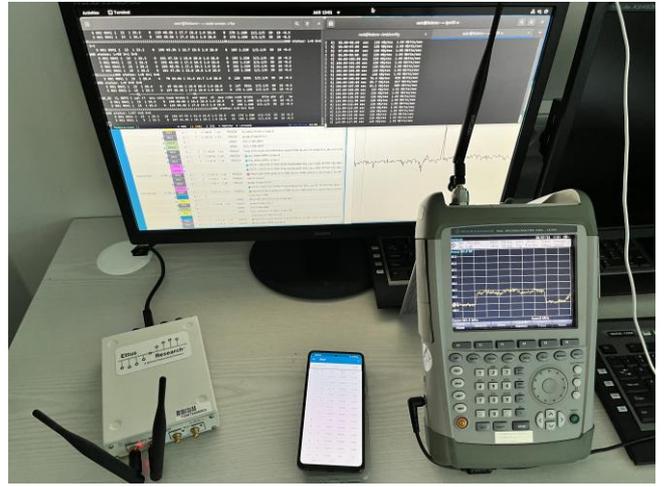

Figure 2. O-RAN testbed setup: Amarisoft CU/DU, USRP B210 O-RU (jammer or O-RU), Samsung A52s 5G UE (legitimate or launching SS), and spectrum analyzer.

Our O-RAN testbed is built on the Amarisoft software platform, which provides CU/DU functionality. The O-RU is realized by attaching a USRP B210 to the Amarisoft CU/DU through the USB 3.0 cable, emulating O-FH. Connected UE (5G smartphone – Samsung A52s) transmits data with the *iperf* program (a tool for data stream generation and performance measurement).

Near-RT RIC enables the deployment of dedicated xApps that protect RAN against jamming and a signaling storm. Its operation is emulated by the *wrapper* to Amarisoft Application Programming Interface (API), i.e., the Python code that acts as an intermediary abstraction layer between xApps in Near-RT RIC and Amarisoft CU/DU, which is an E2 node.

An E2 interface application protocol called E2AP is specified over SCTP/IP as the transport protocol [1][2]. On top of E2AP, application-specific controls and events are conveyed through E2 service models (E2SM). The RIC services, named *REPORT*, *INSERT*, *CONTROL*, *POLICY*, and *QUERY*, are provided on O-CU-CP, O-CU-UP, and O-DU for xApps to access messages and measurements and enable control of these entities from the Near-RT RIC. Different data types are grouped for each RIC service as a so-called *style*.

In our testbed, a dedicated application between Amarisoft CU/DU and xApps, which represents the E2 interface, has been developed based on the

*WebSocket* and utilizes Amarisoft API for the CONTROL and REPORT services. The RIC services (both REPORT and CONTROL) are provided to the xApps. Our Near-RT RIC platform supports the following E2SM styles and data defined by the O-RAN Alliance WG3 [13].

Finally, in our experiments, the transmit power of all devices (legitimate and attacking) was 10 mW, the center frequency in the downlink (DL) was set to 881.5 MHz, in the uplink (UL) to 836.5 MHz, and the system bandwidth to 5 MHz.

## IV. Jamming detection and mitigation xApp

The 5G system is quite robust against interference caused by jammers in the long run. While the jamming signal increases the noise floor at the receiver, the MCS adaptation or retransmissions prevent the link from failing. An update of MCS (after the signal-to-noise ratio change is reported) may take several milliseconds, and similarly, a retransmission delay may not be acceptable for URLLC. In addition, jamming can be easy and inexpensive to launch with low-cost software-defined-radio (SDR) equipment, making it more likely to occur. Utilizing SDR, our testbed was jammed with white noise in a continuous or keyed manner (repeatable sequence of 100 ms of jamming and 100 ms of no jamming). This resulted in increased packet latency (measured via *ping* command). The non-jamming scenario allows 95% of packets to reach the destination and return in less than 40 ms (recall that the utilized testbed is not optimized for latency). The permanent jamming increased this time to 80 ms. The worst performance was observed under keyed jamming, requiring a 200 ms time budget as a result of packet retransmissions. Thus, detecting jamming is crucial for URLLC and is to be carried out in Near-RT RIC for delay minimization.

### A. Jamming detection

In [14], we proposed a jamming detection method based on the analysis of the *Channel State Information* and *Reference Signal Received Power* reports by UE in DL. Here, we present a better-performing method based on the statistics of correct transport block reception messages: the ACK/NACK messages for each transport block in DL and available in E2 nodes. In the experiment, it has been observed that the ACK message statistics were significantly altered by jamming. This can be justified by the anomalous nature of the distortion created by a jammer, not taken into account while designing the radio interface. On the other hand, the typical radio communication effects, such as shadowing or inter-cell interference, are considered for network design, preventing multiple NACK messages from occurring.

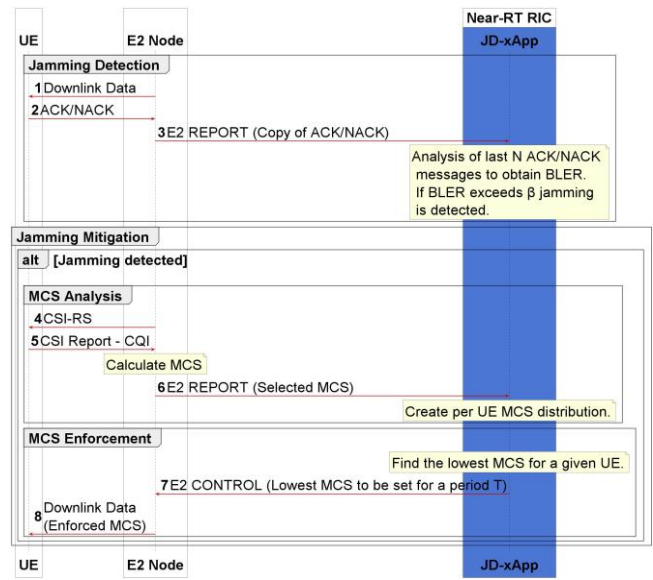

Figure 3. UML sequence diagram for jamming detection and mitigation.

The message exchange between E2 Nodes, UE, and Jamming Detection (and mitigation) xApp (JD-xApp) is shown in Figure 3. First, UE sends ACK/NACK messages via PUCCH (Physical Uplink Control Channel) or PUSCH (Physical Uplink Shared Channel) to E2 nodes. Next, these reports are sent to the JD-xApp over the E2 interface. Several methods can be used to analyze such a dataset for an anomaly. We propose to analyze the mean Block Error Rate (BLER) over the last transmitted $N$ blocks. If BLER is above threshold $\beta$, we assume that jamming occurs, preventing RAN from correct MCS selection.

Note that the long observation window $N$ (high number of ACK/NACK reports) makes the algorithm more robust against false alarms but also less efficient in short-term jamming detection. If $\beta$ is too low, a small increase in mean BLER (caused by varying channel conditions) is detected as jamming. Most importantly, JD-xApp should calibrate $\beta$ and $N$ in each E2 node, as various schedulers can maintain various mean BLER levels. The calibration of $\beta$ and $N$ can be done by collecting ACK/NACK measurements while the network is operating without a jamming signal. The non-jammed state can be confirmed by some other measurements, e.g., using a spectrum analyzer. Based on this dataset, various values of $\beta$ and $N$ can be tested, calculating the probability of a false alarm ($P_{fa}$). For the selected values of $P_{fa}$, multiple ($\beta$, $N$) pairs are available. One should be selected based on the below discussion.

Our measurements show that the longer the observation window $N$, while maintaining such $\beta$ value to keep $P_{fa}$ constant (based on previous calibration), the higher the probability of detection ($P_d$) of jamming. Moreover, the keyed jamming is more challenging to detect. The measurements were carried out for varying UE posi-tions in the radius of 1-15 m. In our indoor environment, the power (transmitted from gNodeB or jammer) was attenuated in the range of 30-55 dB. This mobility dynamics does not allow for performance evaluation for one specific value of Signal-to-Jamming power Ratio (SJR). For jamming received with higher power (e.g., SJR around 0 dB), the negative impact on the 5G link performance is higher, but it is easier to detect. Low-power jamming (e.g., SJR around 20 dB) is difficult to detect but causes hardly any harm to the link quality. In the test scenario, $P_d$ exceeds 50% for both jamming types while $N = 10000$.

The proposed jamming detection requires a very limited set of information, i.e., ACK/NACK messages. Extending its capabilities, e.g., to estimate jammer location, can be possible only if the required information is available at a given O-RAN interface, e.g., UEs location. As for the utilized testbed, the algorithm's extension is limited by the set of reports provided by Amarisoft.

*B. Jamming mitigation*

Mitigation of jamming is a difficult task as straightforward solutions, such as transmit power increase or carrier frequency reconfiguration, are difficult when multiple cells must be coordinated while limiting inter-cell interference and maintaining continuous service provision. On the other hand, a 5G network can adapt to some interference or jamming present in a wireless channel using channel state measurements and MCS selection. However, standard link-adaptation mechanisms can be too slow, especially for a keyed jamming, causing link failures.

In our testbed, it is proposed that the MCS distribution for a given UE be analyzed when the jamming is present and that minimal observed MCS be assigned. While reduced, MCS can result in higher resource utilization, decreasing the number of UEs to be served in parallel; if the proposed scheme is not employed, potentially all the URLLC devices connected can have too high transmission latency, preventing their application from being performed successfully.

The JD-xApp must subscribe to the DL MCS values assigned to a given UE by the E2 node. Its distribution over some time is estimated within JD-xApp. If jamming is detected, JD-xApp sends a proper policy via the E2 interface, preventing the scheduler from assigning the jammed UE higher MCS than the one specified in the policy.

In our experiment, the MCS of transmission to jammed UE was set to 1 to obtain maximal protection against jamming. The jamming mitigation performance can be assessed by observing packet round trip time $\tau$ (using the *ping* command). The

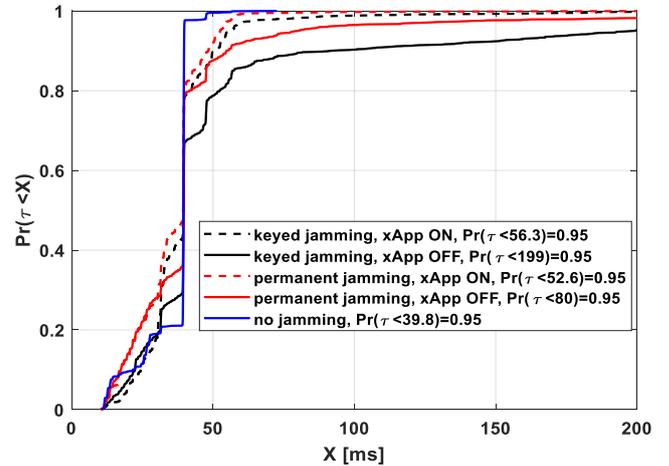

Figure 4. CDF of packet delay (by ping tool) with/without jamming and with/without jamming mitigation xApp.

Cumulative Distribution Function (CDF) of the packet delay in the absence and the presence of two types of jamming, as well as in the case when JD-xApp is applied, is shown in Figure 4. One can observe that jamming causes a *heavy tail* of this distribution if some transport blocks are received with an error requiring retransmission. Jamming detection and mitigation xApp allows for reducing the time budget required for 95% of packets to around 56 ms for both permanent and keyed jamming. This is a nearly 35% reduction, from 80 ms to 52.6 ms, in the case of permanent jamming, and nearly fourfold reduction, from 199 ms to 56.3 ms in the case of keyed jamming. While there is a slight difference between both jamming types when mitigation xApp is on, this seems to be statistically negligible. This proves that our jamming mitigation xApp is highly effective in our testbed environment. Both jamming detection and mitigation should work effectively for various jammer distance and transmission power configurations as long as the observed SJR is within the useful range of all MCSs.

**V. SS detection and mitigation xApp**

O-RAN architecture allows for the protection of resources caused by SSs in the access channels of an mMTC network. Typically, SS can be launched by hacked mMTC devices. The connected devices usually have fixed locations and can be identified during initial access based on their Timing Advance (TA) parameter, i.e., propagation delay time between UE and gNodeB. For the TA parameter to be useful, the devices must be stationary, e.g., as in the network of sensors. This allows selective blocking of a malicious device that constantly sends connection requests aiming at flooding the Core Network (CN) control plane. Thus, SS detection is automatically the mitigation technique, i.e., the detected source of the attack is consequently eliminated.

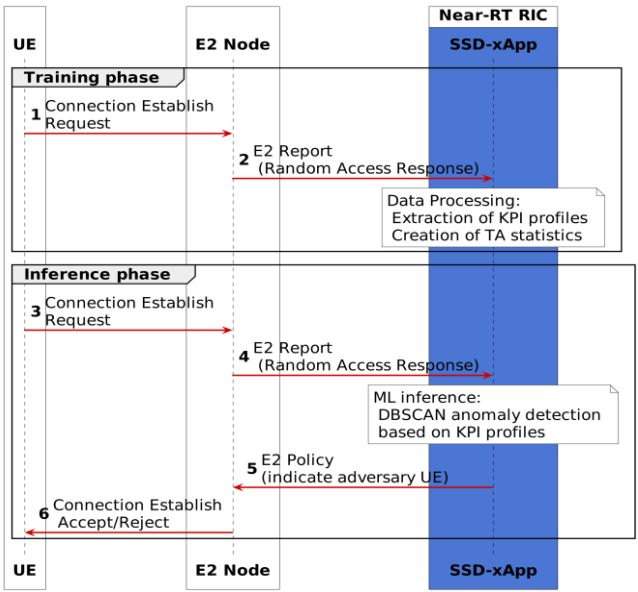

Figure 5. UML sequence diagram for signaling storm detection and mitigation.

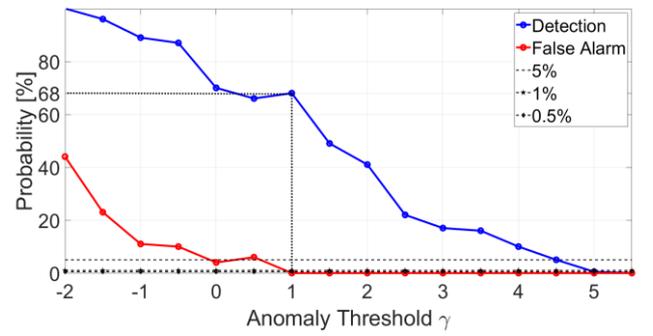

Figure 6. Probability of detection $P_d$ and false alarm $P_{fa}$ vs. anomaly threshold $\gamma$.

In [15], we have proposed a dedicated SS detection xApp (SSD-xApp) for this application scenario. The SSD-xApp is placed in a Near-RT RIC and utilizes O-RAN interfaces to capture and analyze control plane messages exchanged between gNodeB (or an E2 node) and UE to protect CN resources by rejecting connection requests from the signaling storm attack. The related UML sequence diagram is depicted in Figure 5.

First, the SSD-xApp must build a Key Performance Indicator (KPI) profile that contains, per TA statistics, the mean and standard deviation of the number of UE connection attempts observed during consecutive time intervals for each TA. Whenever UE sends a connection-establish request, the Random Access Response (RAR) using the E2 interface is transferred from gNodeB to SSD-xApp to create the KPI profile.

After the training phase, the SSD-xApp goes to the inference phase. Again, it utilizes the E2 interface to capture RAR from the E2 Node, but now, it uses the KPI profile to detect potential threats. The so-called anomaly value is calculated as the (estimated) *standard score* of the number of connection requests associated with a particular TA for a given period, i.e., the currently observed number of connection requests minus the mean divided by the standard deviation of the number of connection requests estimated in the training phase (and available in the KPI profile). If the calculated anomaly value exceeds the threshold $\gamma$ (maximal allowed estimated *standard score*) for a particular TA, the SSD xApp formulates a policy to reject connection-establish requests from UEs of this TA. Note that the same TA may be reported for several UEs, and applying security policy may occasionally result in the rejection of some legitimate requests.

Building on the idea presented in [15] (simulated with 100 mMTC devices and 5 adversaries), we tested the proposed SSD-xApp in the real-world O-RAN testbed described above. One smartphone was used as a legitimate stationery mMTC device that sent connection-establish requests three times per minute. Another smartphone emulated the adversary mMTC device, launching 15 signaling storm attacks per hour; one attack composed of 12 connection establish requests sent in 2 s intervals. The first part of the SSD-xApp evaluation aimed to create the KPI profile. It was built in one hour and contained the average number and the standard deviation of connection-establish requests sent by the legitimate mMTC device within one minute.

The critical observation after obtaining the measurement-based KPI profile is that even if there is only one legitimate stationary mMTC device, the KPI profile contains statistics for three TAs. This reflects the properties of a network, where TA varies over time. After obtaining a reliable KPI profile, we examined the SSD-xApp running under different anomaly threshold $\gamma$. The SSD-xApp continuously tracks the connection-establish requests from the last minute, calculates anomaly value, and performs SS detection. Each evaluation was 1 hour long to provide statistically reliable results. The results are depicted in Figure 6. Note that the selection of anomaly threshold $\gamma$ is crucial for balancing the probability of detection ($P_d$), the probability of false alarm ($P_{fa}$), and the probability of the rejection of legitimate UE along with an attacker. In the case of mMTC traffic, keeping low $P_{fa}$ and not deteriorating transmission quality is crucial. Thus, $P_d$ can be "sacrificed" (selected lower) within a reasonable range.

It can be seen in Figure 6 that the SSD-xApp can keep $P_d$ around 68% while reducing $P_{fa}$ below 0.5% for $\gamma = 1$. As in the case of mMTC traffic, low $P_{fa}$ is of high importance, and the $P_d$ at the level of 68% is a good result, i.e., it allows the detection and blocking of most of the SS attacks. On the other hand, it is beneficial not to disturb network operation when less intensive signaling storms occur, as they will not cause access denial.

## VI. Conclusions

An O-RAN platform with its xApps running in Near-RT RIC can be used to monitor, analyze, and eliminate threats providing a higher level of security. As demonstrated, dedicated xApps can be developed to protect the network against anomalies such as jamming or signaling storms in the example URLLC and mMTC use cases, although their application is not limited to these examined scenarios. The presented jamming detection and mitigation xApp, in the URLLC scenario, allows for reducing the required packet latency level for 95% of packets for both permanent and keyed jamming. The SS detection and mitigation xApp in the mMTC scenario results in a sufficient probability of detection while reducing the probability of false alarms below 0.5%.

A challenge for future research is to develop JD-xApp for smart jamming and a more sophisticated jamming signal, e.g., a synchronization signal. It should also involve a quantitative evaluation of JD with respect to the signal-to-jamming power ratio. Moreover, the utilization of early-stage registration messages to detect SS caused by the moving adversary and the impact of E2 message delay on the SSD-xApp performance are topics for prospective studies.

**Acknowledgment**

This work was funded by the National Centre for Research and Development in Poland within the 5GStar project CYBERSECIDENT/487845/IV/NCBR/2021 on "Advanced methods and techniques for identification and counteracting cyber-attacks on 5G access network and applications".

**Literature**

[1] O-RAN WG1, "O-RAN Architecture Description," June 2024, https://orandownloadsweb.azurewebsites.net/specifications
[2] M. Hoffmann, *et al.*, "Open RAN xApps Design and Evaluation: Lessons Learnt and Identified Challenges," *IEEE J. Selec. Areas Comm.*, vol. 42, no. 2, pp. 473-486, Feb. 2024
[3] M. Polese, L. Bonati, S. D'Oro, S. Basagni, T. Melodia, "Understanding O-RAN: Architecture, Interfaces, Algorithms, Security, and Research Challenges," *IEEE Comm. Surveys & Tutorials*, 2023, Vol. 25, No. 2, pp. 1376-1411
[4] M. Liyanagea et al., "Open RAN Security: Challenges and Opportunities," *J. Netw. Comput. Appl.* 214, C (May 2023)
[5] J. Green et al. "Implementing and Evaluating Security in O-RAN: Interfaces, Intelligence, and Platforms," *IEEE Network*, 29 July 2024
[6] Open RAN MoU Signatories: "Open RAN Technical Priorities Focus on Security," 2023.
[7] National Security Agency and Cybersecurity and Infrastructure Security Agency (USA) "Open Radio Access Network Security Considerations", September 15th, 2022.
[8] O-RAN Alliance WG 11, "O-RAN security threat modeling and remediation analysis 6.0," O-RAN.WG11.O-RAN-Threat-Model-v04.00, 2023
[9] J. Cao et al., "A Survey on Security Aspects for 3GPP 5G Networks," *IEEE Commun. Surv. Tut*. vol. 22, no. 1, pp. 170-195, Firstquarter 2020
[10] Y. Arjoune, S. Faruque, "Smart Jamming Attacks in 5G New Radio: A Review," Annual Comp. and Comm. Workshop and Conference, Las Vegas, NV, USA, 2020, pp. 1010-1015.
[11] A. Tabiban, H. A. Alameddine, M. A. Salahuddin, R. Boutaba, "Signaling Storm in O-RAN: Challenges and Research Opportunities," *IEEE Comm. Mag.*, vol. 62, no. 6, pp. 58-64, June 2024
[12] O-RAN Alliance, "Zero Trust Architecture for Secure O-RAN," white paper, May 2024
[13] O-RAN Alliance WG3, "O-RAN E2 Service Model (E2SM), RAN Control 6.0" O-RAN.WG3.E2SM-RC-R003-v06.00, 2024
[14] P. Kryszkiewicz, M. Hoffmann "Open RAN for detection of a jamming attack in a 5G network" IEEE VTC Spring 2023
[15] M. Hoffmann, P. Kryszkiewicz, "Signaling Storm Detection in IIoT Network based on the Open RAN Architecture," IEEE INFOCOM 2023 Workshops, Hoboken, NJ, USA, 2023, pp. 1-2.

**Hanna Bogucka** is a professor at the Institute of Radiocommunications at PUT and the co-founder of RIMEDO Labs. She is involved in research in wireless cognitive and green communication. She is a member of the Polish Academy of Sciences. She also serves as the Member-at-Large of the IEEE Communications Society BoG.

**Marcin Hoffmann** is a Senior R&D engineer at Rimedo Labs working on O-RAN software development solutions. He is also a Ph.D. candidate at PUT. His research include utilizing machine learning for 5G/6G network management.

**Pawel Kryszkiewicz** is an associate professor at the Institute of Radiocommunications, PUT. He is also the co-founder and Technical Director of RIMEDO Labs. He is a Fulbright Alumnus of the Worcester Polytechnic Institute, MA, and the author of over 100 publications on multicarrier systems, dynamic spectrum access, and O-RAN.

**Łukasz Kułacz** is an assistant professor at the Institute of Radiocommunications. His main fields of interest are Open RAN implementations and radio resource management.